\documentstyle[aps,prl,epsf,twocolumn]{revtex}
\begin{document}
\twocolumn[\hsize\textwidth\columnwidth\hsize\csname
@twocolumnfalse\endcsname
\title{Periodic instanton method and macroscopic quantum tunneling between two weakly linked Bose-Einstein condensates}
\author{Yunbo Zhang$^{1,2}$ and  H.J.W. M\"uller-Kirsten$^{1}$}
\address{$^1$Department of Physics, University of Kaiserslautern,
 D-67653 Kaiserslautern, Germany \\
$^2$Department of Physics and
 Institute of Theoretical Physics,
 Shanxi University, Taiyuan 030006, P.R. China}
 \maketitle
\begin{abstract}
A new method is used to investigate the tunneling
 between two weakly--linked Bose--Einstein condensates
 confined in double--well potential traps. 
The nonlinear interaction between
 the atoms in each well contributes
 to a finite chemical potential, which,
 with consideration of periodic
 instantons, leads to a 
remarkably high tunneling frequency.
 This result can be used to 
interpret the newly found Macroscopic
 Quantum Self Trapping (MQST) effect.
 Also a new kind of first--order crossover
 between different regions is predicted.
\end{abstract}
\vspace{0.2cm}
]

Following the first observation of Bose--Einstein condensation (BEC)
 in dilute gases of trapped alkali atoms, remarkable progress has
 been made both theoretically
 and experimentally \cite{1}. In particular,
 interference between two freely expanding 
condensates has been observed after
 switching off the double--well potential
 that confines them\cite{2}. By
 using a thinner barrier between
 the two condensates it should be possible to
establish reliably a weak link and study quantum 
tunneling, or the Josephson effect,
 for atoms. Aspects of these questions have 
already been studied theoretically 
in the limit
 of noninteracting atoms \cite{3} 
and for small--amplitude Josephson 
oscillations \cite{4,5}. 

Here we develop
 another theoretical method for
 a sensitive and precise
 investigation of the tunneling between 
two condensates. The almost trivially looking 
problem of the tunneling behavior in 
a double--well potential has attracted
 much attention from theorists for
 decades. For a single particle, the solution can be
 found even in quantum mechanics textbooks\cite{6}. 
The advantage of a nonperturbative method,
 as presented here, is that it gives not only
 a more accurate description of the tunneling 
phenomena but also a comprehensive
 physical understanding in the context
 of quantum field theory. The periodic instanton 
configurations, which have been shown to be a useful tool
in several areas of research 
such as spin tunneling\cite{7}, 
bubble nucleation\cite{8}
 and gauge field theory\cite{9}, enable also the investigation of
 the finite temperature behavior of these systems.
 In the case of the Bose--Einstein system, however, we need to 
evaluate the tunneling frequency for a finite
 chemical potential even at zero temperature,
 due to the nonlinear interaction between
 the confined atoms. Therefore the chemical 
potential here replaces the position 
of the excited energy and gives 
rise to an expected higher tunneling frequency. 

A novel nonlinear effect has been
 predicted to occur in the
 Bose--Josephson Junction(BJJ)\cite{10}:
 The self--trapping of a BEC population
 imbalance arises because of the interatomic
 nonlinear interaction in the Bose gas.
 This was considered to be a novel ``macroscopic quantum 
self--trapping'' (MQST) and was predicted
 to be observable under certain
 experimental conditions. The three parameters,
 i.e. the ground state energy $E^0$, the
 interaction energy $U$, and more 
importantly, the tunneling amplitude $K$, 
are still undetermined for a specific
 geometry of the trap
 and have been taken as constants in refs.\cite{10,11}.
 Here we present a rigorous derivation 
of these quantities and find that they
actually depend
on the number of atoms $N$. This $N$--dependence 
refines the conclusions and makes the 
self--trapping easier to observe.

The macroscopic wave function $\Phi$ associated
 with the ground state of a dilute Bose gas
 confined in the potential $V_{ext}(r)$ obeys
 the well--known Gross--Pitaevskii Equation (GPE), 
which can be obtained using a variational
 procedure, i.e. 
 $i \hbar \partial \Phi/\partial t 
= \delta E/\delta \Phi ^{*}$. 
The energy functional $E$ is defined by
\begin{equation}
\label{e}
E[\Phi] =\int d^3r\left[ \frac{\hbar ^2}{2m}
\left| \nabla \Phi \right|^2+V_{ext}(r)
\left| \Phi \right|^2+\frac{g}2 \left| \Phi \right|^4\right] 
\end{equation}
where $g=4\pi \hbar^2a /m$ is the
 interatomic coupling 
constant with $a$ the 
$s-$wave scattering length. The three terms in 
the integral are the kinetic energy
 of the condensate $E_{kin}$, the (an)harmonic 
potential energy $E_{ho}$, and the
 mean--field interaction 
energy $E_{int}$, respectively.
 In the simplest case of an isotropic 
harmonic trap $V_{ext}(r)=m\omega _0^2r^2/2$, 
these energies, which in the Thomas-Fermi approximation
(TFA) assume the simple values 
\begin{equation}
\frac{E_{kin}}N=0,\quad \frac{E_{ho}}N=\frac 37\mu _{TF},\quad 
\frac{E_{int}}N=\frac 27\mu _{TF},
\end{equation}
can be calculated beyond the TFA\cite{1,4,12} as
\begin{eqnarray}
\frac{E_{kin}}N=\frac 52 C,
 \frac{E_{ho}}N=\frac 37\mu _{TF}+ C,
\frac{E_{int}}N=\frac 27\mu _{TF}-C
\end{eqnarray}
where $C=\frac{\hbar ^2}{mR^2}\ln \left( \frac R{1.3a_{ho}}\right)$ is a 
correction term due to the presence of a boundary layer near the condensate 
surface. Here $N$ is the number of atoms  and 
the harmonic oscillator 
length $a_{ho}=(\hbar /m \omega_0)^{1/2}$ is 
introduced for simplicity.
Correspondingly the chemical potential in the TFA
$\mu_{TF}=\frac{\hbar \omega_0}{2}
\left(15Na/a_{ho}\right)^{2/5}$, related to the radius of the condensate
 $R$ through $\mu_{TF}=m\omega _0^2R^2/2$, is modified beyond the TFA as
$\mu=\mu_{TF}+3C/2$.
 We note that in the derivation,
 the wave function is normalized to $N$. If one
 uses instead a wave function normalized to unity,
 the following correspondence should be realized 
\begin{eqnarray}
U_{1,2}N_{1,2} \rightarrow 2\frac{E_{int}}N,
 \qquad E_{1,2}^0 \rightarrow \frac{E_{kin}}N+\frac{E_{ho}}N
\end{eqnarray}
and we obtain the ground state 
energy $E_{1,2}^0$ and the
 interaction self energy $U_{1,2}N_{1,2}$ for
 the isolated traps
 with $N_1=N_2=N$ as in ref.\cite{10}
\begin{eqnarray}
E_{1,2}^0 =\frac 37\mu _{TF}+\frac 72 C, \qquad 
U_{1,2}N_{1,2} =\frac 47\mu _{TF}-2 C
\end{eqnarray}
Considering a condensate of $N=5000$ sodium 
atoms confined in a symmetric spherical 
trap with frequency $\omega _0=100Hz$, 
we have $E^0 =1.18n{\rm K}, UN=1.03n{\rm K}$, quite close
 to the values estimated in \cite{10}. 

\textit{Calculation of the tunneling
 frequency by means of the periodic
 instanton method:} We study 
the amplitude for tunneling  between
 the two condensates confined in the wells 
of an external double--well potential
\begin{equation}
\label{dw}
V_{ext}(x)=\frac{m\omega _0^2}{8x_0^2}(x^2-x_0^2)^2
\end{equation}
The two minima are located at $\pm x_0$ on the $x$--axis, and 
the harmonic oscillation frequency near these minima is $\omega_0$. 
The barrier height between the two wells $V_0=\frac 18 m\omega _0^2 x_0^2$ 
is assumed to be high enough so that the overlap between the wave functions 
relative to the two traps occurs only in the classically forbidden region 
where interaction can be ignored and one can safely use the WKB wave 
function approximately\cite{4}. The tunneling amplitude $K$ in ref. 
\cite{10} can be calculated by different methods, and we demonstrate 
in this work the use of the nonperturbative instanton approach.
 It is easily shown that this tunneling amplitude is just the quantity 
${\cal R}$ of \cite{11} (up to a minus sign), if
 one observes the orthogonality property
 of the eigenfunctions $\int dx \Phi _i^{*}(x)\Phi _j(x)=\delta _{ij},
 i,j=1,2$, with $\Phi_{1,2}(x)$ the local modes in each well, 
which are taken as the harmonic oscillator single
 particle ground state wave function in ref. \cite{11}. The
 nonlinear interaction between the atoms 
in the same well will be included,
 which modifies only the chemical potential
 $\mu $ to or beyond the TFA.

Now we turn to the field theory description
 of the GPE. To this end we consider
 a scalar field in a 1--dimensional time plus
 1--dimensional space. After
 a Wick's rotation $t =-i \tau$ the
 Euclidean--Lagrangian equation of motion
 for a finite chemical potential takes the form
\begin{equation}
  \frac 12 m\left( \frac{dx}{d\tau}\right) ^2-V_{ext}(x)=-\mu \label {mo}   
\end{equation}
The reason why we can handle a nonlinear
 problem by means of a linear equation of motion
 is that we discuss the tunneling behavior
 in the barrier region where the nonlinear 
interaction is negligible. However,
 there are obvious differences between
 the BEC tunneling system and the usual
 one-body problem, i.e. the nonlinear
 interaction contributes a finite
 chemical potential, which is just the
 integration constant on the right hand 
side of eq.(\ref{mo}). The classical turning
 points on both sides of the barrier
 can be determined by the relation
 $V(x_{1,2})=\mu$ as suggested
 in ref. \cite{4}. For a noninteracting 
system the chemical potential approaches 
the ground state energy corresponding 
to the vacuum instanton case in \cite{14}.

Solving this Euclidean time classical equation
 in the usual way \cite{14}
 one obtains the periodic instanton 
solution in terms of the Jacobian
 elliptic function $x_c=2x_0kb(k)/\omega _0
 \mathop{\rm sn} \left( b(k)\tau \right)$
 with the parameters defined as 
\begin{equation}
b(k)=\frac{\omega _0}2
\sqrt{\frac 2{1+k^2}},\qquad k^2=\frac{1-u}{1+u},\qquad u=\sqrt{\frac \mu {V_0}}
\end{equation}
The Euclidean
 action for this solution in half of the imaginary period $T=2{\cal K}(k)/b(k)$ can be
 obtained through $S=\int_{-T/2}^{T/2}d \tau 
\left( \frac 12 m(dx/d\tau)^2+V_{ext}(x)\right)=W+\mu T/2$ with
\begin{equation}
W =\frac 23\frac{8V_0}{\omega _0}(1+u)^{1/2}\left( {\cal E}(k)-u{\cal K}(k)\right) 
\end{equation}
where ${\cal K}(k)$ and ${\cal E}(k)$ are
 complete elliptic integrals of
 the first and second kinds with
 modulus $k$, respectively. The 
frequency of tunneling between the two condensates 
is then given by the energy level splitting
 of the two lowest states,
i.e.  $\Omega =\Delta E/\hbar=2K/\hbar=2{\cal R}/\hbar$ 
and can be calculated by
 means of the path integral method as\cite{14}
\begin{equation}
\label{fr}
\Omega =\frac 1 \hbar Ae^{-W/\hbar }=\frac{\sqrt{1+u}}{2 {\cal K}(k^{\prime })}\omega _0\exp \left[-W/\hbar \right]
\end{equation}
We emphasize here that this 
formula has been proven to
 be valid for the entire
 region when the chemical
 potential is below the barrier height. The
 condition $V_0=\mu$ determines
 the sphaleron configuration, where
 a type of
phase transition may occur. In the TFA this means 
\begin{equation}
x_{0}=2R=2a_{ho}\left( \frac{15N_Ta}{2a_{ho}}\right) ^{1/5}
\end{equation}
where $N_T=N_1+N_2$ is the total 
 number of atoms in both wells together.
 Therefore for a specific type of trapped 
atoms and a given double--well potential
 with separation $x_0$ (atom number $N_T$) there
 exists a critical number of atoms $N_{c1}$ (critical separation 
$x_{c1}$) determined by the above equation,
 below (above) which the tunneling 
process will give the main
 contribution to the tunneling
 amplitude. However, above this
 critical number of atoms or below this critical
 separation value another process, i.e.
 the over--barrier activation will dominate
 (which is definitely 
  not ``thermal activation'' as
 in spin tunneling since the
 temperature is zero)(cf. Fig. 1). Between these
 two processes there exists
 a crossover. A more explicit 
condition for this critical
number of atoms (separation between 
the two minima) can be derived beyond the TFA:
\begin{equation}
x_{0}=2R\sqrt{1+\frac 35 \left( \frac{15N_Ta}{2a_{ho}}\right) ^{-4/5}\ln \left( \frac{15N_Ta}{2a_{ho}1.3^5}\right)}
\end{equation}
As an example, we consider two
 weakly--linked condensates
 of $N_T=10^4$ sodium atoms, confined
 in two symmetric spherical 
traps with frequency $\omega _0=100Hz$ as 
in ref. \cite{10}. The critical
 value for $x_{c1}$ in the TFA is $x_{c1}=24.58\mu m$
 or more accurately beyond the TFA $x_{c1}=25.29\mu m$.
 We note that here the height of the potential barrier
 is $V_0 =2.21n{\rm K}$ and the ground 
state is located at $\hbar \omega_0/2=0.38n{\rm K}$ so
 that there are several energy levels
 beneath the barrier height. This means
 the interaction between the atoms contributes
 to the chemical potential, which effectively
 raises the classical turning points
 to a remarkably high level. Although
 the atoms remain in the ground state,
 the interaction
 energy is so strong 
that the vacuum instanton method can no longer be applied.
 We have to resort to the periodic instanton method, as will be shown below.

\begin{figure}[ht]
\epsfxsize=2.5 in\centerline{\epsffile{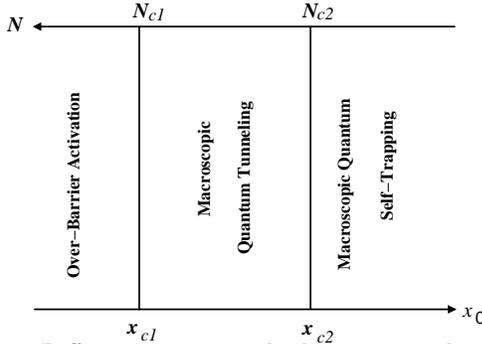}}
\caption{Different regions with the corresponding critical parameter}
\end{figure} 

\textit{Low--energy limit:}
We now consider
 the ``low-energy'' limit, $\mu \rightarrow 0$.
 As in the case of a uniform 
Bose gas, the number of atoms in
 the ground state can be
 macroscopic, i.e., of the order of the total number in one potential
 well, when the chemical potential becomes equal to
 the energy of the lowest state, which, in our
 1--dimensional case here, is $\mu \rightarrow 
\mu _c=\frac 12\hbar \omega_0$. 
The lower boundary for the chemical
 potential in fact implies
 that $R_c=a_{ho}$, i.e. the radius 
of the condensate should
 never be less than 
the harmonic oscillator
 length $a_{ho}$. We thus
 have a result similar to that in the vacuum
 instanton case\cite{14}
 and the ``low energy'' 
limit here is only meaningful
 in this sense. Expanding eq. (\ref{fr}) 
far below the barrier height, i.e., around
 the modulus $k \rightarrow 1$,  or 
equivalently evaluating the
 tunneling amplitude in the vacuum 
instanton method \cite{14}, we obtain
 for the tunneling frequency 
\begin{equation}
\label{vacuum}
  \Omega=2\sqrt{\frac{6S_c}{\pi \hbar }}\omega _0\exp \left( -S_c/\hbar \right)
\end{equation}
with the Euclidean action
\begin{equation}
  \frac{S_c}{\hbar}=\frac 23\frac{8V_0}{\hbar \omega _0}=\frac 23\frac{x_0^2}{a_{ho}^2}
\end{equation}
This result can be compared with that of ref. \cite{11}
 where the problem of tunneling
 in the BEC system is considered. 
We reexpress the tunneling frequency 
of ref. \cite{11} in the notation
 of the harmonic oscillator length $a_{ho}$ as
\begin{equation}
\label{mb}
\Omega=\frac{x_0^2}{a_{ho}^2}\omega _0e^{-x_0^2/a_{ho}^2}
\end{equation}
which, however,  gives not only a smaller exponential 
contribution $8V_0/\hbar \omega _0$ (there is a $2/3$ factor)
 but also an inaccurate
 prefactor $x_0^2\omega _0/2\triangle=\omega _0S_c/\hbar$.
 The source of this inaccuracy is the adoption 
of the too simple harmonic oscillator 
wave function of a single particle, which obviously
 oversimplifies the Bose--Einstein
 condensation tunneling problem. At least one should
 use the WKB wave function in the tunneling
 region, and it can be shown that this corresponds to the vacuum
 instanton result we present here. For
 the agreement between WKB and vacuum 
instanton methods we refer to ref. \cite{15}. 

\textit{Observation of Macroscopic Quantum Self Trapping:}
 The periodic instanton result
 will lead to a rapidly growing
 behavior for the tunneling
 frequency \cite{7} when the chemical potential,
 i.e. the number of atoms, is increased.
 According to ref. \cite{10}, for
 a fixed value of the initial population
 imbalance $z(0)$ and phase difference $\phi(0)$, if
 the parameter $\Lambda$ exceeds
 a critical value  $\Lambda_c$, the population
 becomes macroscopically self--trapped with 
a nonzero average population
 difference $N_1-N_2$. There are different
 ways in which this state can be achieved, and 
all of them correspond to the so-termed MQST condition that
\begin{equation}
  \label{MQST}
  \Lambda=\frac{UN_T}{2K}>\Lambda_c=2\left(
\frac{\sqrt{1-z(0)^2} \cos \phi(0)+1}{z(0)^2}\right)
\end{equation}
This requirement is actually that the modulus
 of the elliptic function, which appears in the population
 oscillation solution, should be
 larger than 1 so that
 the elliptic function $cn$ will be
 replaced by $dn$ and the oscillation period 
is shortened from $8k{\cal K}(k)/C \Lambda$ 
to $4 {\cal K}(1/k)/C \Lambda$.
 The parameters $UN_T$ and $K$ are taken
 as constants in ref. \cite{10}. Considering the
 fact that they are actually $N$--dependent as
 in our calculation above, we can refine
 the conclusions of refs. \cite{10,11}. To access
 the region of self--trapping,
 that is, $\Lambda >\Lambda_c$, it is better
 to lower the value of $K$ by making a higher 
barrier height $V_0$ through increasing
 the separation $x_0$ or the
 oscillation frequency $\omega_0$, 
than to increase the number of atoms as
 suggested in ref. \cite{10}. In fact,
 the quantity $UN_T$ here is proportional
 to $\mu _{TF}\sim N^{2/5}$ which means that
 increasing the number of atoms
 will not increase the interaction
 energy significantly, and at
 the same time the tunneling amplitude will be
 increased more drastically.
 Thus, contrary to the result of
 refs. \cite{10,11}, we find that
 the MQST will occur when the number of atoms is smaller 
(instead of larger) than a critical value $N_{c2}$, i.e.
 we should decrease the number of
 atoms instead of increasing it (Fig. 1).
 Inserting the values of the interaction energy
 and the tunneling amplitude
 into eq.(\ref{MQST}) we can obtain this critical number of
atoms for a given potential geometry.
 The parameters which can be adjusted are the  number
of atoms $N_T$, the oscillation frequency $\omega _0$,
 and the separation distance
 between the two condensates $x_0$. Fig 1. shows
 the three different regions for different
  numbers of atoms and distances between the potential wells.
 When $x_0$ ($N_T$) is smaller (larger) than
 the critical value $x_{2c}$($N_{c2}$),
 the atoms will oscillate between these 
two potential wells. Once we increase
 the separation above (or decrease the number of atoms below)
 this critical value, the MQST will occur, i.e., most of the
atoms will tend to remain in 
their appropriate
 wells, leading to only a small
 oscillation around a fixed population difference. 

We take the initial condition for the population 
difference to be $z(0)=0.4$ and the zero--phase case $\phi (0)=0$
as an example. Other cases with, for example, a non--zero phase difference
 give rise to only a different critical parameter $\Lambda_c$. For sodium atoms
 confined in the double--well potential with $\omega_0=100Hz$,
 we show numerically in Fig. 2 
the critical line between the three different regions in the TFA and beyond it. 
The upper region marks the self--trapping region,
 the lower the over--barrier activation.
 Quantum tunneling occurs only for a small range of the parameter.
In the experiment \cite{2}, the barrier was generated by an off-resonance 
(blue detuned) laser beam. To make our results more applicable to experiment 
we denote on the right vertical axis the corresponding barrier height in 
units of nK. We also find that the tunneling
 will be suppressed when the separation or 
the number of atoms satisfies $x_0>28\mu m$ or $N_T>12500$
(in the TFA $x_0>26\mu m$ or $N_T>12000$). The crossover will occur directly 
between the self--trapping and the over--barrier regions, quite similar 
to the first--order transition in spin tunneling\cite{7}.

\begin{figure}[ht]
\epsfxsize=2.5 in\centerline{\epsffile{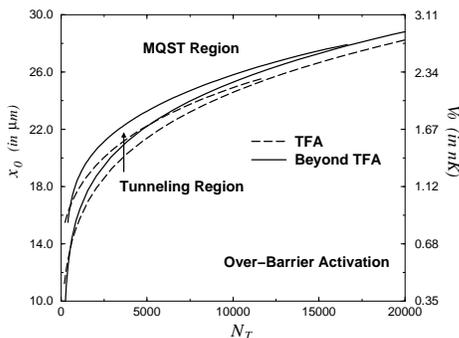}}
\caption{Critical line for MQST effect. Solid line: results beyond the TFA 
where the
parameters take the values $UN_T=8/7 \mu_{TF}-4C$ and $\mu=\mu_{TF}+3C/2$.
Dashed line: the TFA results where
$UN_T=8/7 \mu_{TF}$ and $\mu=\mu_{TF}$ are used in the numerical 
simulation. }
\end{figure} 

In conclusion we can say, we have shown that 
the periodic instanton method can be used
 to investigate the tunneling problem in BEC systems
 at zero temperature. The tunneling amplitude and the
 nonlinear interaction energy between the atoms can 
be calculated analytically beyond the TFA. 
The MQST is more easily observed if one
takes into account the $N$--dependence of the 
tunneling amplitude $K$ and the self interaction energy $UN_T$.
 The crossover between the different
 regions may be of the first--order type
 when the two minima of the potential wells are separated 
sufficiently far or the
number of confined atoms in the potential well is large enough.

Y. Z. thanks Prof. F. Dalfovo for
 stimulating his interest in this subject
 and acknowledges support by an Alexander von Humboldt Foundation
Fellowship.
 Helpful discussions with Dr. Xiao--Bing Wang, Dr. Wei--Dong Li
 and Prof. J.--Q. Liang are appreciated.

\end{document}